\documentclass[12pt]{article}
\usepackage{latexsym}
\usepackage{graphics}

\usepackage {amsfonts}
\usepackage{epsfig}
\def\thefootnote{\arabic{footnote}}

\def\p{\partial}

\newcommand{\be}{\begin{equation}}
\newcommand{\ee}{\end{equation}}
\def\bea{\begin{eqnarray}}
\def\eea{\end{eqnarray}}
\def\Tr{{\rm Tr}}

\newcommand{\eref}[1]{(\ref{#1})}

\def\IR{\mathbb{R}}
\def\IZ{\mathbb{Z}}
\newcommand{\bR}[0]{{\bf R}}
\newcommand{\SO}{\mathrm{SO}}

\newcommand{\CO}{{\cal O}}

\newcommand{\ads}[1]{{\rm AdS}_{#1}}

\newcommand{\sph}[1]{{\rm S}^{#1}}
\font\mybb=msbm10 at 10pt
\def\bb#1{\hbox{\mybb#1}}
\newcommand{\fso}{\mathfrak{so}}
\def\bR {\bb{R}}
\def\bZ {\bb{Z}}

\begin{document}

\rightline{hep-th/0407212} \rightline{ITFA-2004-30}

\vskip 1.5 cm
\renewcommand{\thefootnote}{\fnsymbol{footnote}}
\centerline{\Large \bf Some Aspects of the AdS/CFT Correspondence}

\vskip 1.5 cm \centerline{{\bf Jan de Boer\footnote{\tt
jdeboer@science.uva.nl}, Liat Maoz\footnote{\tt
lmaoz@science.uva.nl} and Asad Naqvi\footnote{\tt
anaqvi@science.uva.nl} }}

\vskip .5 cm \centerline{\it Instituut voor Theoretische Fysica}
\centerline{\it Valckenierstraat 65, 1018XE Amsterdam, The
Netherlands} \vskip .5 cm

\setcounter{footnote}{0}
\renewcommand{\thefootnote}{\arabic{footnote}}

\vskip 1.5 cm
\begin{abstract}

This is a very brief review of some aspects of the AdS/CFT
correspondence with an emphasis on the role of the topology of the
boundary and the meaning of the sum over bulk geometries. To
appear in the proceedings of the 73rd Meeting between Physicists
and Mathematicians ``(A)dS/CFT correspondence,'' Strasbourg,
September 11-13, 2003.

\end{abstract}

\newpage

\section{Introduction}

Since its incarnation in 1997 \cite{ads1,ads2,ads3} (for a review
see \cite{ads4}), the AdS/CFT correspondence has been one of the
prime subjects of interest in string theory. It provides a duality
between a theory with quantum gravity in $d$ dimensions and a
field theory in $d-1$ dimensions. This is a rare example where we
have a complete non-perturbative definition of string theory in a
certain background, and quite amazingly it is equivalent to just
an ordinary field theory. Strongly coupled string theory is
equivalent to weakly coupled field theory and vice versa, and
therefore the AdS/CFT correspondence can help understand the
physics of strongly coupled gauge theories. At the same time,
although Anti-de Sitter space has different asymptotics than
Minkowski space or our universe (as far as we know), the
properties of gravity at short distances should be somewhat
independent of the  asymptotic
behavior of the space. Therefore the AdS/CFT correspondence should also be
useful in understanding the puzzles associated to
quantum gravity, in particular those associated with black hole
creation and evaporation and information loss in black holes.

Though it is easy to say these words, to actually implement them
in practice is not quite so straightforward. Given a certain
manifold $M$ on which the CFT lives, which can be either
Lorentzian or Euclidean, the dual gravitational description
involves a sum over all geometries whose conformal boundaries are
equal to $M$. This by itself is a mathematical question, namely
the classification of all solutions to the Einstein equations with
a negative cosmological constant with given conformal boundary.
However, this is not the full story, since we should really sum
over all solutions of the string theory equations of motion with
the right asymptotic behavior. String theory has various other
fields in addition to the metric, and therefore this set can be
larger than the set of purely gravitational solutions.
Furthermore, the string theory configurations that have to be summed over  do
not need to be weakly coupled everywhere and  can in principle
include stringy objects such as branes in the interior. The full
classification of all such solutions is still in its infancy and a
further understanding seems crucial in order to make progress in
our understanding of quantum gravity.

With these motivations, we will here
briefly summarize some of the known problems and solutions
associated to finding the bulk solutions with a priori given
asymptotia. We have no pretense of being complete and/or
exhaustive, the main purpose of these notes is to provide some
food for further thought.

There are in principle many cases to consider, and we organized
them as follows. In section~2 we consider solutions with Euclidean
signature and a single connected boundary. In section~3 we
consider static solutions with Lorentzian signature and a single
connected boundary. In section~4 we discuss solutions with more
than one boundary, and in section~5 we briefly comment on the
relation of all this with the puzzles associated to singularities
and black holes in quantum gravity. In section~6 we comment on the
role of Chern-Simons theory in $2+1$ dimensions; this is an
especially interesting case since gravity has no propagating
degrees of freedom in $2+1$ dimensions. Finally, in section~7 we
mention some examples that involve time-dependent geometries, and
in section~8 some of the problems associated to extending all this
to zero and positive cosmological constant are summarized.

\section{Euclidean, single boundary setups}
\paragraph{AdS/CFT -- the statement}
The original statement of AdS/CFT is a relationship between the
partition function of string theory on AdS$\times X$ geometry and
that of a CFT living on the boundary of AdS:
\begin{equation}
Z_{CFT}(\partial M;\gamma) = \int {\cal D} \Phi_{\rm string}
e^{-S_{SFT}}\stackrel{\stackrel{{\rm saddle}}{  {\rm
point}}}{\approx} \sum_i Z_{\rm String}(M_i). \label{partition}
\end{equation}
The CFT lives on $\partial M$ which carries a metric in a fixed
conformal class indicated by $\gamma$, and the LHS of
(\ref{partition}) is the CFT partition function. We have
schematically written the string path integral as an integral over
string fields approaching $\partial M$ on the boundary, and
$S_{SFT}$ represents the string field theory action. In going from
this somewhat schematic expression to a more useful expression
$\sum_i Z_{\rm String}(M_i)$, we have made a saddle point
approximation.  $M_i$ are backgrounds which satisfy the string
equations of motion  and have $(\partial M;\gamma)$ as their
conformal boundary. In most cases, the string coupling constant
and the inverse curvature radius of the AdS space are free
parameters and we can take them arbitrarily small. In this
limit\footnote{On the dual CFT side, the interpretation of this
limit depends on the model, but generically involves taking a
strong coupling limit, and also often involves sending the rank of
a gauge group to infinity} we can replace the string theory
partition function with the classical supergravity contribution,
which is simply $Z_{\rm Sugra}=e^{-S(M_i)}$ where $S(M_i)$ is the
classical supergravity action evaluated on the solution $M_i$.

Notice the sum over different manifolds with the same conformal
boundary. This plays a crucial role in the study of phase
transitions in the boundary CFT. In a certain regime of
parameters, generically one of the spaces $M_i$ will dominate the
sum on the RHS of (\ref{partition}). However, by varying the
parameters, which of these $M_i$'s dominates may change, leading
to phase transitions. The existence of a phase transition is
perhaps surprising if the boundary $\partial M$ is compact (as
will mostly be the case in what we study below). However, in the
large $N$ limit, we can still have sharp phase transitions, even
on a compact volume.

An important ill-understood feature of (\ref{partition}) is the
precise relative normalization of the contributions on the right
hand side. In the absence of a background independent formulation
of string field theory is is not obvious how to compute these from
first principles. The naive guess to take just the supergravity
actions is incorrect in the example discussed in \cite{Farey}, but
the microscopic origin of the relative normalization found in that
paper is not clear.

In this section, we will discuss Euclidean setups, with the bulk
space $M$ being a solution to Einstein's equations with $\Lambda
<0$, which has a single boundary $\partial M$. We restrict
ourselves to the case of a four dimensional bulk which is one of the best studied cases \cite{anderson}.

\subsection{$\p M=S^3,~~~M=B^4$}
This is the best understood example. Euclidean $\ads{4}$ can be
described as the open unit ball $B^4$, with coordinates $x_i$ such
that $\sum_{i=1}^4 x_i^2 < 1$ and the metric
\be
ds^2= {4 dx^2   \over (1-|x|^2)^2}.
\ee
This metric does not extend to the boundary at $|x|^2=1$. However,
the metric can be extended to the boundary by defining a function
$f$ on the closure of $M$ such that it has a simple zero at the
boundary and is positive in the interior. Then, the metric
$d\tilde{s}^2=f^2 ds^2$ extends to a metric on the boundary, but
given a bulk metric, only the conformal structure of the boundary
metric can be uniquely determined. Following a theorem of Graham
and Lee,  for every conformal structure on $\sph{3}$ sufficiently close to the
standard one, there exists
a metric on $B^4$ with that $S^3$ as a conformal boundary.
Analogous statements hold for the scalar fields as well as the
gauge fields- bulk fields are uniquely specified by their behavior
on the boundary. This means that when we apply the AdS/CFT
correspondence in this background, we can compute the correlation
functions in the boundary theory by evaluating the bulk action for
field configurations which asymptotically approach a given
boundary data. On the RHS of \eref{partition}, there is only one
term in the summation.

\subsection{$\p M =S^1 \times S^2, ~~~ M=S^1 \times R^3 ~~{\rm and}~~ M=R^2 \times S^2$}
With the boundary $S^1_\beta \times S^2$ (where $S^1_{\beta}$ is a circle of radius $\beta$),  there are two known
asymptotically AdS bulk solutions with this boundary. One is AdS
itself (with topology $S^1 \times R^3$), with the Euclidean time
direction being a circle. This background is appropriate
for the finite temperature bulk physics at temperature $1/\beta$. Another solution with the
same boundary behavior is Euclidean AdS-Schwarzschild (this has
topology $R^2 \times S^2$). The boundary theory is a CFT at finite temperature.

It was shown by Witten \cite{phasetrans} that the first solution dominates the partition function computation at
low temperatures while the latter becomes dominant at high temperatures. This difference in behavior corresponds
to the confinement-deconfinement phase transition in the field theory. This phase transition was first studied by
Hawking and Page \cite{hawkingpage} who showed that above a critical temperature, thermal radiation is unstable to the formation of a
AdS Schwarzchild black hole. There are, in fact, two black hole solutions, with different masses for a given value
of $\beta$ (the temperature). The smaller value of the masses leads to a black hole with a negative specific heat,
which means that the black hole is unstable to decay as Hawking radiation. For the higher value of the masses, the
Hawking radiation is in thermal equilibrium with the thermal radiation in the background.

\subsection{$\p M=T^3, ~~~M=R^2 \times T^2$}
These are the AdS toroidal black hole metrics, which are of the
form
\be
ds^2=U^{-2} dr^2 +U^2 d\theta^2 + r^2 ds_{T^2}^2,
\ee
where $U^2=r^2-{2m \over r}$.
The conformal boundary of this space is $S^1_\theta \times
T^2=T^3$. Given a boundary metric, there are actually infinitely
many different ways of filling in the bulk metric, each
corresponding to a choice of one cycle in $T^3$ which is 'filled
in' to obtain a bulk solution. For details, see \cite{anderson}.
The boundary theory is a finite temperature CFT on $T^2$. The
multiple classical solutions perhaps correspond to different
phases in this theory. Heuristically, the CFT partition function
can be written as \be Z_{\rm CFT} \sim \sum_{g \in SL(3,\IZ)/H}
\exp\Bigl(-I(M_g)\Bigr). \ee This expression should be taken
with a large grain of salt. We do not really understand how to
perform this sum here. In one lower dimension, for a two
dimensional boundary, a similar summation was performed in
\cite{Farey} where the elliptic genus of the conformal field
theory was computed by writing it as a sum over different
asymptotically $\ads{3} \times \sph{3}$ bulk geometries (also see
section \ref{mathur}). The $\ads{3}$ string theory should reduce
to a Chern Simons theory at large distances. The calculation of
the bulk partition function in this Chern-Simons theory is a state
in the space of conformal blocks of the boundary theory and
therefore transforms non-trivially under the modular group
\cite{gukovetal}. On the other hand, the string theory computation
in \cite{Farey} gives a modular invariant partition function. This
apparent paradox is resolved \cite{gukovetal} by the special
appearance of a modular invariance restoring chiral ``spectator
boson'' on the boundary.

\subsection{AdS Taub-bolt metrics}
The AdS Taub-bolt metrics are locally asymptotically AdS. The conformal boundary is an $S^1$ bundle over $S^2$,
with non-zero first Chern number. For vanishing first Chern number, the boundary is the product space $S^1 \times
S^2$ and the space is asymptotically AdS. This is one of the cases we discussed above. However, for non-vanishing
first Chern number, $k$, the conformal boundary is a squashed $S^3$ with $|k|$ points identified along the $S^1$.
These metrics have a U(1) isometry which acts on the $S^1$ fiber in the natural way. For the AdS Taub-Bolt
metric, the fixed point set of this isometry is two-dimensional (called a bolt). The line element is given by \be
ds^2=-{3 \over 4\Lambda}E\Bigl[ {F(r)\over E(r^2-1)} (d\tau +E^{1/2}\cos\theta d\phi)^2 +{4 (r^2-1) \over
F(r)}dr^2+(r^2-1)(d\theta^2+\sin^2\theta d\phi^2) \Bigr], \label{adstb} \ee with \be F_{\rm
Bolt}(r)=Er^4+(4-6E)r^2+\Bigl(-Es^3+(6E-4)s+{3E-4 \over s}\Bigr)r +4-3E, \ee and \be E={2ks-4 \over 3(s^2-1)}. \ee
Here $\Lambda <0$ is the cosmological constant, $\tau$ has period $\beta={4\pi E^{1/2} \over k}$,
 $s$ is an
 arbitrary parameter (the bolt is at $r=s$) and $k$ is the Chern number of the $S^1$ bundle over $S^2$, which is the conformal boundary of this solution. $|k|$ points on the $S^1$ fiber are identified.

There is another class of closely related metrics for which the fixed point set of ${\p \over \p \tau}$ is just a
point. These are the AdS Taub-Nut metrics. The line element for these metrics has the same form as \eref{adstb}
but the function $F(r)$ is now given by \be F_{\rm NUT}(r)=Er^2+(4-6E)r^2+(8E-8)r+4-3E. \ee Now, $E$ is an
arbitrary parameter which parameterizes the squashing of the $S^3$ which is the conformal boundary.

The AdS TN and AdS TB have the same asymptotic behavior for $k=1$. For $|k| > 1$, if we identify $|k|$ points on
the $S^1$ fiber of the AdS TN solution, we obtain a space which has the same boundary structure as the AdS TB
solution with parameter $k$. This identified AdS TN solution, however, has a conical singularity at the origin,
which can be smoothed out.

For computation for the analogue of the ADM mass, and action for these solutions, we need to compare it to some
reference metric which has the same boundary behavior. The reference metric is taken to be the AdS Taub-Nut
metrics discussed above, with appropriate identifications along the $S^1$ fiber to get the same asymptotic
structure. Then, the Hamiltonian calculation reveals that there are two AdS Taub-Bolt metrics with the same
temperature, but different masses. The one with the lower masses is thermodynamically unstable, since it has a
negative specific heat. In addition, as in the AdS case, there is a phase transition in the system (for $k=1$).
The AdS Taub-Nut solution exists for all temperatures. However, the AdS Taub-Bolt solution can only exist for
temperatures above a minimum value $T_0$.

Furthermore, since we have multiple bulk solutions with the same
boundary behavior, the partition function of the boundary CFT will
receive contributions from the different bulk spaces with the same
boundary behavior. To determine which one dominates, we need to
evaluate the action for these solutions. It can be shown
\cite{hawking} that for temperatures below  $T_1~~~ (> T_0)$, the
AdS Taub-Nut background is favored, whereas for temperatures above
$T_1$, the AdS Taub-Bolt solution dominates, and the AdS Taub-Nut
background will decay into it. This presumably corresponds to a confinement/deconfinement phase transition
for the boundary theory living on the squashed $\sph{3}$.

\section{Static Lorentzian spacetimes, $\Lambda < 0$}
In the previous section, we discussed Euclidean situations, where
specifying the boundary values of the various fields at the
conformal boundary determines the bulk configurations, in some
cases uniquely, and in others, upto a few discrete choices. The
situation in Lorentzian signature is more subtle.  The
normalizable mode solutions to the equations of motion, which
exist in Lorentzian signature, can be arbitrarily added to a bulk
solution with a given boundary behavior without affecting the
boundary behavior. The choice of the normalizable part of the
solution corresponds to the choice of state in the conformal field
theory in which the partition function and hence the correlation
functions are computed. Here again we will only deal with the case of four
dimensional bulk, as it is one of the most studied cases \cite{AndLor}.

\subsection{$\p M = \IR \times S^2, ~~~ M= \IR\times \IR^3=\IR^4$}
\label{usual}
This is the usual Lorentzian AdS/CFT setup. The boundary CFT lives
on $\IR \times S^2$.
Given a certain boundary metric (with non-negative Ricci scalar),
a bulk metric always exists with that boundary behavior \cite{AndLor}. The
uniqueness of such a metric is not guaranteed in general. However
for the boundary metric $ds^2 = -dt^2 + ds^2_{S^2}$, there is a
unique globally static bulk metric with conformal compactifiable
smooth acausal equal time slices. This is just the standard metric
of $\ads{4}$  \cite{AndLor}.

\subsection{$\p M = \IR \times S^2, ~~~ M=\IR \times (\IR^+\times S^2) $}

Looking now at the same boundary manifold but at $M=\IR \times (\IR^+\times S^2)$ as a bulk manifold, we find a
rather different situation. Taking the boundary metric to be again $ds^2=-dt^2 + ds^2_{S^2}$, one can check that
the following family of 1-parameter bulk metrics all have the required asymptotics - these are the Lorentzian AdS
Schwarzschild black holes with the metric \be ds^2=-U^2 dt^2 +U^{-2} dr^2 +r^2 (d\theta^2 +\cos^2 \theta d\phi^2)
\label{adsblackhole}, \ee where \be U^2=1+r^2 -{2m \over r} \,\,\,,\,\,\, m>0.\ee This background corresponds to a
thermal state in the boundary CFT. It differs from the usual setup in (\ref{usual}) by the choice of the state on the boundary.

One can also show that these AdS Schwarzschild black hole metrics
are the unique globally static metrics smooth up to the horizon
with conformal compactifiable smooth acausal equal time slices
\cite{AndLor}.

The dual CFT correspondning to the boundary conditions of both the global $AdS_4$ metric described in section 3.1 and
these AdS Schwarzschild black holes is a CFT defined on a spatial
manifold $S^2$. To discuss this theory at finite temperature, one effectively needs to calculate the partition function on
$S^2\times S^1$, where the radius of the extra $S^1$ factor is related to the inverse of the temperature, and it can be
thought of as the time direction, Wick rotated to Euclidean signature. The calculation of the partition function then
follows the one we had in section 2, for Euclidean spacetimes. Therefore the Hawking-Page phase transition occurs here
and is seen in the Field theory as a confinement - deconfinement transition. Of course in cases where the field theory is
not in a finite temperature, it is hard to tell which geometry would dominate the partition function.

\subsection{$\p M = \IR \times T^2, ~~~ M= \IR \times (D^2 \times S^1) $}

Let us look at the boundary metric $ds^2 = -dt^2 + ds^2_{T^2}$.
Then one can show \cite{AndLor,Galletal} that all the globally static
metrics on $M$ with such asymptotics and with conformal
compactifiable smooth acausal equal time slices are of the ``AdS soliton''
type discussed by Horowitz and Myers \cite{HorMyers}:  \be ds^2 = -r^2dt^2
+ U^{-2} dr^2 +U^2d\phi^2 +r^2d\theta^2, \ee where \be U^2 =
r^2-{2m \over r} \,\,\,,\,\,\, m>0.\ee  $\phi$ is a periodic angle of
period
$\beta=\frac{4\pi}{3(2m)^{1/3}}$, and $\theta$ is of arbitrary
period.

One can also show that for any given
boundary metric on $T^2$ there are countably many such filling
metrics parameterized by the choice of an $S^1=\p D^2$.

These geometries have the interesting property that their mass is negative (relative to the choice where conformal
flatness means zero energy). This fact has a natural interpretation on the CFT side - it was shown in
\cite{HorMyers} that the corresponding CFT has a negative Casimir energy, related to the breaking of supersymmetry
on the CFT by the boundary conditions on the fermions. In fact it was conjectured in \cite{HorMyers} that the ``AdS
soliton'' metrics are the lowest energy solutions with these given boundary conditions.

\subsection{$\p M = \IR \times T^2, ~~~ M= \IR \times (\IR^+ \times T^2) $}

In this case, taking again the boundary metric to be $ds^2 = -dt^2 +
ds^2_{T^2}$, one can show that there is a 1-parameter
family, this time of toroidal black holes with these asymptotics.
This family of toroidal Kottler metrics is given by \be ds^2 =
-U^2dt^2 + U^{-2}dr^2 +r^2d\phi^2 +r^2d\theta^2, \ee where as before \be
U^2 =r^2 - {2m \over r}\,\,\,,\,\,\, m>0, \ee and where both $\phi,\theta$
are periodic of arbitrary period.
These metrics have a horizon at $r^4= 2m$ which is $\IR\times T^2$. As
before
these filling metrics are the unique ones which are globally
static and with conformal compactifiable smooth acausal equal time
slices \cite{AndLor}.

The energy of these black holes is greater than that of the AdS solitons of the same boundary structure, in
accordance with the conjecture made by Horowitz and Myers \cite{HorMyers}. However, a thermodynamical analysis
\cite{SSW, ThDyn,Vanzo}, shows that the free energy of these black holes can be greater or smaller than that of
the solitons, leading to a phase transition, somewhat similar to the Hawking-Page transition we mentioned earlier.
It has been shown that small, hot black holes are unstable and decay to small, hot solitons. Large cold black
holes are stable. The order parameter for the transition depends both on the horizon area and on the temperature
of the black hole (which are two independent parameters for these black holes). On the side of the CFT, this phase
transition can be related to a confinement - deconfinement transition \cite{SSW, Page}.

\section{Multiple boundary configurations}

In cases where the boundary of spacetime has multiple disconnected components, the issue of a dual holographic
description is more involved. On the one hand the holographic theory is defined on a union of disjoint manifolds.
There is no obvious way in which the theories on the different manifolds are coupled, and apriori it seems natural
to expect that the holographic theory would just be the product of the theories on each one of the boundary
components. On the other hand the bulk theory seems to induce correlations between the different boundary regions.

This seeming puzzle bears a somewhat different nature depending on whether one is discussing Euclidean or
Lorentzian settings.

In the Lorentzian case, for asymptotically AdS spacetimes (i.e. $\Lambda<0$), a {\it topological censorship
theorem} was proved \cite{TCT}, which basically states that under certain conditions, the presence of multiple boundaries
forces the bulk to be separated by horizons, in such a way that different boundary components are not causally connected
through the bulk. This
implies that the different holographic theories living on the different boundary components would indeed be
uncorrelated and will not interact dynamically. The only correlations could be ones in initial states of the
theory \footnote{One example for this is the case of Schwarzschild AdS black holes, and in particular the BTZ black
hole. These were studied in \cite{eternal} and we would make a few comments about them below.}. Let us state the
topological censorship theorem more precisely now:
Let ${\cal M}'$ be a globally hyperbolic spacetime with boundary, with timelike boundary ${\cal I}$ that satisfies
the average null energy condition \footnote{The average null energy condition states that for
each point $p$ in ${\cal M}$ near ${\cal I}$ and any future complete null geodesic $s\to\eta(s)$ in ${\cal M}$
starting at $p$ with tangent X, $\int_0^\infty Ric(X,X)\,ds \geq 0$ ($Ric(X,X)$ denotes $R_{ab}X^aX^b$) .This condition is
satisfied by spacetimes created from physically reasonable matter sources.
}. Let ${\cal I}_0$ be a connected component of ${\cal I}$ of ${\cal M}'$.
Furthermore assume that either (i) ${\cal I}_0$ admits a compact spacelike cut or (ii) ${\cal M}'$ satisfies the
generic condition \footnote{A spacetime satisfies the generic condition if every
timelike or null geodesic with tangent vector X contains a point at which $X^aX^bX_{[c}R_{d]ab[e}X_{f]}$ is
nonzero.}. Then ${\cal I}_0$ cannot communicate with any other component of ${\cal I}$, i.e. $J^+({\cal
I}_0)\bigcap({\cal I}\setminus {\cal I}_0)=\emptyset$.

In the Euclidean case, for asymptotically AdS Einstein spacetimes, the puzzle is avoided due to a theorem by
Witten and Yau \cite{WY} , basically stating that if one of the boundary components has $R>0$, then the boundary is
connected. This theorem was later generalized by Cai and Galloway \cite{CG} to cases where the boundary has zero scalar
curvature. Let us state the general theorem: Let $M^{n+1}$ be a complete Riemannian manifold which admits a
conformal compactification, with conformal boundary $N^n$, and with the Ricci tensor of $M$ satisfying $Ric \geq
-ng$ such that $Ric \to -ng$ sufficiently fast on approach to conformal infinity \footnote{ i.e. $r^{-2}(Ric+ng)\to 0$ as
$r\to 0$ where the bulk metric is expanded in a neighborhood of the boundary as $g=\frac{1}{r^2}(dr^2 + g_r)$, and the
conformal boundary is at $r\to 0$. For some discussion on a physical interpretation of these conditions, see \cite{Brett}.}. If
$N$ has a component of
nonnegative curvature, then the following holds: (i) $N$ is connected (ii) If $M$ is orientable, then
$H_n(\bar{M},Z)=0$ (iii) The map $i_*:\Pi_1(N)\to\Pi_1(\bar{M})$ ($i$=inclusion) is onto.

This theorem therefore implies that the puzzle we described does not arise in asymptotically AdS Einstein
spacetimes with nonnegative boundary curvature. One might wonder then about the case where the boundary has
negative curvature. In such cases, it can be shown that the holographic theory living on the negative curvature
boundary would be unstable for any boundary dimension $n\geq 3$. However, the Witten Yau theorem can be
avoided if we turn on extra supergravity fields, and thus look not at Einstein spacetimes, but rather at spacetimes obeying
the more general supergravity equations. The instability related to negative curvature boundary can also be avoided if we
look at specific settings of 3-dimensional bulk (i.e. $n=2$). Such examples will be presented in section 4.3.

Let us now discuss a few examples of multi boundary situations where one can say something about the AdS/CFT
correspondence. We'll focus in the case where the bulk is 3-dimensional, where things are better known. In fact in
the case of 3-dimensions, the only solution to Einstein's equations with a negative cosmological constant is
locally $AdS_3$. Different spacetimes can only differ from each other by global identifications.

Starting from Lorentzian configurations, one can build configurations with multiple boundary components by taking
a 2-dimensional slice of global $AdS_3$ , cutting and gluing it along geodesics and then letting it evolve in time
\cite{CutGlue} . Such constructions lead, in agreement with the topological censorship theorem, to spacetimes with the
same number $h$ of boundaries and horizons, and with any number $g$ of handles behind the horizons.

In the special case where there are only two boundaries ($h=2\;,\;g=0$), the spacetime describes the eternal BTZ black hole.
\subsection{Eternal BTZ}

\label{eternalbtz}
\begin{figure}
  \begin{center}
    \epsfysize=1.5in
    \mbox{\epsfbox{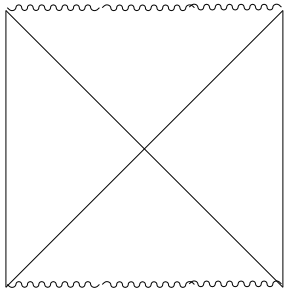}}
  \end{center}
 \caption{Penrose diagram of the eternal BTZ}
\label{btzpenrose}
\end{figure}

An eternal BTZ black-hole has two boundaries.
More precisely, it has two asymptotically $\ads{3}$ regions each of which is separated from the other by a horizon (see the Penrose diagram in fig \ref{btzpenrose}).  Locally, it is isometric to $\ads{3}$ but differs from it in its global properties. Three dimensional anti-de Sitter space is a maximally symmetric space of constant negative curvature. It is the hyperboloid
\begin{eqnarray}
    \ads{3}  &\hookrightarrow& \bR^{2,2} \nonumber \\
    -u^2 - v^2 & +& x^2 + y^2 = -l^2  ,
 \label{embedding}
\end{eqnarray}
in flat $\bR^{2,2}$. By construction, the isometry group is $\SO(2,2)$. The Killing vectors of the metric generate the Lie algebra  $\fso(2,2)$ of the isometry group, and
are described in terms of the embedding space $\bR^{2,2}$ as
\begin{equation}
  J_{ab}=x_b \partial_a -x_a \partial_b ~,
\label{bulksl2gens1}
\end{equation}
with $x^a \equiv (u,v,x,y)$ and $x_a=\eta_{ab}\,x^b$, with
$\eta_{ab}=(-\,,-\,,+\,,+)$.

The BTZ black holes are obtained by identifying $\ads{3}$ by
the discrete action generated by the Killing vector \cite{BHTZ}
\begin{equation}
  \xi_{\rm BTZ}={r_+ \over l}J_{12}-{r_- \over l}J_{03}-J_{13}+J_{23}.
\label{nonextident}
\end{equation}
In the non-extremal case, $r_+^2 -r_-^2 >0$ and by a $\SO(2,2)$
transformation, $\xi_{\rm BTZ}$ can be brought into the form:
\begin{equation}
  \xi_{\rm BTZ}^\prime={r_+ \over l}J_{12}-{r_- \over l}J_{03}.
\label{xiprime}
\end{equation}
The mass and angular momentum of the black hole are given by
\begin{equation}
  M={1 \over l^2}(r_+^2+r_-^2)\quad , \quad J={2 \over l} r_+r_-.
\end{equation}
The extremal black hole is obtained by taking the limit $r_+ \to r_-$ in (\ref{nonextident}), so that  the generator becomes
\begin{equation}
 \xi_{BTZ} \to \frac{r_+}{l}\left(J_{12} - J_{03}\right) - J_{13} + J_{23}~.
\end{equation}
The holographic description of the space-time is given in terms of two non-interacting identical CFTs. In Lorentzian AdS/CFT, a holographic description includes the specification of a state in the CFT. The relevant state in two CFTs which describes the BTZ background is a particular entangled state given by
\begin{equation}
| \Psi \rangle = \sum_n e^{-{\beta \over 2}E_n} | E_n \rangle_1 \times | E_n \rangle_2,
\end{equation}
where $|E_n\rangle_{1,2}$ denotes an energy eigenstate in the two CFTs and $\beta={\pi l \over r_{+}}$.

Computation of the correlation functions involving only operators in one of the two CFTs in this entangled state lead to thermal correlation functions:
\be
\langle \Psi | \CO_1 | \Psi \rangle = \sum_n e^{-\beta E_n} {}_1\langle E_n | \CO_1| E_n \rangle_1
=\Tr \Bigl(\rho_\beta \CO_1 \Bigr),
\ee
where $\rho_\beta$ is the thermal density matrix.

We will come back to the eternal BTZ in Section \ref{open}.
\subsection{Two null cylinder boundaries}
This space has two boundaries, each of which are null cylinders, {\em i.e.} flat space with a compact null direction. The metric is  given by
\begin{equation}
ds^2=l^2\left(-(dt)^2+(d\phi)^2+2 \sinh(2z) dt d\phi + (dz)^2\right) \, ,
 \label{eq:fmetric}
\end{equation}
with $\phi$ being an angular coordinate taking values in $[0,2\pi)$, and the coordinates $(z,\phi,t)$ give a global parameterization of the space. The two boundaries are at $z \rightarrow \infty$. Unlike the BTZ, the boundaries are not separated by a horizon. The space is stationary but not static. It is a quotient of $\ads{3}$ by action of a subgroup of $\SO(2,2)$ isomorphic to $\bZ$.
\begin{equation}
  {P} \rightarrow e^{t\xi} {P}~~~,~~~t=0,
  \pm 2\pi, \pm 4 \pi, \cdots \quad \forall\,{P}\in \ads{3}
 \label{orbact}
\end{equation}
where
\begin{equation}
  \xi = {1\over 2}\left(J_{02} + J_{13}\right)~.
 \label{eq:killvect}
\end{equation}
This generator is a linear combination of a boost in the ux-plane and
vy-plane in the embedding space $\bR^{2,2}$.

The metric (\ref{eq:fmetric}) is an $S^1$ fibration
over $\ads2$. Indeed, we can rewrite this metric as
\begin{equation}
  g = l^2\left(-\cosh^2(2z)\,dt^2 + dz^2 + \left(d\phi + \sinh (2z)\,dt\right)^2\right)~.
 \label{eq:fibration}
\end{equation}
Compactifying on $\phi$ now gives the metric
  \begin{eqnarray*}
    g_2 &=& -\cosh^2 2z \, dt^2 + dz^2~, \\
    A_1 &= & \sinh 2z \,dt.
\end{eqnarray*}
The metric is precisely that of $\ads{2}$, but there is also a constant electric field.\footnote{The field is constant in the sense that the field strength of the U(1) connection is proportional to the $\ads{2}$ volume form.}

The fact the the boundaries are null cylinders implies that the boundary theory is defined through a discrete light cone quantization procedure. The exact map between the bulk and the boundary theory is still somewhat mysterious.
For more details, see \cite{bns}.

\subsection{Wormholes}
Now looking at Euclidean setups, in order to describe multiple boundary spaces (i.e. wormholes), we must look at
configurations which do not satisfy the conditions of the Witten-Yau theorem. One possibility is to look at spaces
which have two boundaries, each being a Riemann surface of genus $g\geq 2$ , $\Sigma_g$ \cite{WHs}:
\begin{equation}
\label{Riembdys} ds^2=d\rho^2+\cosh^2\rho \,ds^2_{\Sigma_g}.\end{equation} The boundaries have constant negative
curvature, so the Witten-Yau theorem does not apply here, but unlike in higher dimensions, the two-dimensional
field theories on these Riemann surfaces are well-defined and stable. Such spaces can be created by a quotient of
$H_3$ by a discrete subgroup $\Gamma\in SL(2,C)$. The two Riemann surfaces have the same genus, but may differ in
their moduli $t^\alpha$. Performing a quotient of $SL(2,C)$ by a Fuchsian subgroup $\Gamma$ results in two Riemann
surfaces which have the same moduli. Performing a quotient by a quasi-Fuchsian subgroup results in two Riemann
surfaces of different moduli. In fact, according to the Bers simultaneous uniformization theorem \cite{Bers} the
quasi-fuchsian space of a Riemann surface $\Sigma_{g,n}$ of genus $g$ and $n$ punctures: ${\cal QF}(\Sigma_{g,n})$
is homeomorphic to pairs of points in the Teichmuller space of $\Sigma_{g,n}$: $Teich(\Sigma_{g,n})\times
\overline{Teich(\Sigma_{g,n})}$ .

In such cases, the puzzle we described previously for multiple boundary spaces is apparent, and it is not clear
whether correlations would or would not exist between the two boundary holographic theories \cite{WHs}. Actually it is
interesting to note that it is also possible to construct by discrete quotients of $H_3$ , spaces with a single
boundary which could be any Riemann surface $\Sigma_{n,g}$ (this is guaranteed by the classical retrosection
theorem). In this case the discrete subgroup is a Schottky group, and such constructions were described in
\cite{Krasnov}. It would be interesting to see if the geometry with disconnected boundaries (i.e. a union of two such
spaces, each with a single Riemann surface as a boundary) has larger or smaller action than the geometry that
connects the two boundaries. If the disconnected geometry is the dominant one, then it is possible that
correlations between the two boundaries will indeed be suppressed.

Another way to avoid the Witten-Yau theorem and still build a Euclidean spacetime with multiple boundaries of
positive curvature, is to add to the pure Einstein gravity some gauge fields. An example in 4 dimensions was built
in \cite{WHs} , where an $SU(2)$ gauge field was introduced, and the 4-dimensional action is
\begin{equation}
\label{fourdS}
 S \sim \int
d^4x \sqrt{g}[-R+\Lambda+F_{\mu\nu}^aF^{a\mu\nu}].
\end{equation}
 Here the field strength is given in terms of the gauge field by
$F=dA+A^2$ and the cosmological constant is normalized to be $\Lambda=-6$. \footnote{It was shown in \cite{WHs} that this
action is a consistent reduction of 11-dimensional supergravity.} The following two-boundary solution for this
action was constructed:
 \begin{eqnarray}
ds^2 &=& d\rho^2 + e^{2w}ds^2_{S^3} \;\;\; ; \;\;\; e^{2w}={1\over 2}(\sqrt{5}\cosh 2\rho -1)\nonumber \\  A &=&
{1\over 2}\omega^a, \label{Solfourd}
\end{eqnarray}
where $\omega^a$ are the left-invariant, $su(2)$-valued
one-forms on $S^3$, such that $ds^2_{S^3}={1\over 4}\omega^a\omega^a$. For large $\rho$ the geometry becomes that
of $H_4$ and the boundaries are 3-spheres, and for $\rho=0$ we have a finite throat size.
The precise way in which the AdS/CFT correspondence would work here is quite mysterious. It would be very
interesting to understand its exact meaning and interpretation for such configurations.
For more details, other examples of wormhole setups, and some speculations regarding possible resolutions of the AdS/CFT
puzzle for such configurations see \cite{WHs}.

\section{Some interesting questions}
\label{open}
\subsection{ The information paradox}
Hawking showed that black holes are not really black but in fact
emit a thermal radiation. This follows from a semi-classical
analysis. We now imagine matter in a pure state falling into the
black hole. Hawking's semi-classical analysis reveals that it will
eventually be radiated out as thermal radiation, which is a mixed
state. This exposes a paradox: in quantum mechanics,  pure states
evolve to pure states. How then can a pure state evolve into a
mixed state in a black hole background. The information about the
initial pure state seems to have been lost inside the black hole.
This has been called the information loss paradox
\cite{hawkinginfo} --  after matter falls into the black hole, the
correlators with infalling matter decay exponentially,  so if we
wait a long enough time, the correlation functions will eventually
vanish. This violates unitarity.  In AdS, this poses a
particularly sharp paradox since aymptotically AdS black holes can
live for ever.

In section \ref{eternalbtz} we described the eternal BTZ black
hole and its description in the two boundary CFTs in terms of an
entangled state. In \cite{eternal}, Maldacena considered a
deformation of the thermal Schwarzchild AdS state by adding an
operator to the second boundary and showed that the correlations
indeed die off exponentially. This change in the black hole state,
although minor, is still detectable: the one point function of the
same operator in the first CFT which was previously zero is now
non-zero, but dies exponentially fast at a rate $e^{-{ct \over
\beta}}$ where $c$ is a numerical constant. The puzzle is
that correlation functions of the boundary CFT cannot decay at
late times since it signals a loss of unitarity (one can in fact
show that to be consistent with unitarity, the correlations could
be as small as $e^{-cS}$ where $S$ is the entropy of the ensemble
and $c$ is a numerical factor). The resolution comes from the fact
that in AdS/CFT, we need to sum over all geometries with a
prescribed boundary behavior. In fact, there are other geometries
than just the Schwarzchild geometry that we have so far
considered. The additional geometry that provides the relevant
effect consists of two separate global AdS spaces with a gas of
particles on them in an entangled state. This geometry contributes
with a small weight because it has a very small free energy
compared to the Schwarzchild geometry, but it indeed gives a
non-decaying answer of order $e^{-c S}$ as expected from
constraints of unitarity; see also \cite{barbon}, \cite{horomalda}, \cite{solodukhinA}, \cite{solodukhinB}
and \cite{kleban} for further discussion. In the last paper it is
shown that although the sum over geometries does yield a
non-decaying answer, it seems extremely difficult to obtain the
required quasi-periodic answer for correlation functions this way,
unless one manages to perform the sum over geometries in closed
form.

There is another picture of black holes due to Mathur et al that has got some attention recently which we describe in more
detail in section \ref{mathur}. According to this picture, the black hole should be thought of as an ensemble of classical
geometries, each of which has no horizon. The absence of a horizon in each of the different geometries evades the information loss problem.

\subsection{Singularity beyond the horizon}

In the AdS/CFT correspondence, the region outside the horizon of a black hole is represented holographically by a
boundary CFT at finite temperature. Since the black hole singularity is inside the horizon, at first sight it
seems that AdS/CFT cannot be used to gain insight into the nature of the singularity. However, the situation is
more subtle for the eternal black holes, which have multiple asymptotic regions. For example, the eternal BTZ
black hole has two asymptotically AdS regions. The holographic dual is given by two decoupled CFTs living on the
two boundaries, living in an entangled state as described in Section \ref{eternalbtz}. For correlations functions
where all the operators belong to one of the two CFTs, we can trace over the states of the other CFT leading to
correlation functions in a thermal state. Such correlation functions will not contain any non-trivial information
about the physics beyond the event horizon. However, correlation functions of operators in each of the two CFTs
will contain information about the region beyond the horizon. This is most easily seen by using the geodesic
approximation to compute the correlators. For example, for a two point function of operators inserted on the two
boundaries, the WKB approximation is good for bulk fields of large mass. In this approximation,  space-like
geodesics dominate the contribution to the 2-point function, and these geodesics traverse the region behind the
horizon. In the case of the BTZ black hole, these two point computations can be carried out exactly, by using
appropriate bulk to boundary propagators, and then moving the bulk point to the boundary and removing an overall
rescaling. In \cite{KOS}, Kraus et al defined these amplitudes by an analytic continuation procedure from
Euclidean signature.   This analytic continuation can be done in different ways. In one way of performing the
analytic continuation (I) , in Lorentzian signature, the contribution comes from only the region behind the
horizon whereas in the other procedure (II), the contribution comes from both the region outside and behind the
horizon. Since the two analytic continuation procedures are equivalent and finite,  (I) manifestly so, while in
(II), the singularity can be regulated by an $i\epsilon$ prescription inherent in the analytic continuation
procedure, and the contribution from the past and future singularity can be shown to cancel. Its not entirely
clear how much information behind the horizon can really be inferred from this procedure: the fact that we can
obtain the same correlation function by integrating in the region outside the horizon seems to suggest that no
real information behind the horizon can really be contained in these correlations functions. Similar analysis can
be carried out for rotating BTZ's \cite{leviross,vijaylevi}.

The situation for AdS Schwarzchild black holes in higher dimensions is more involved. As was shown in
\cite{fidkowski}, the Penrose diagram is not a square, which results in a contribution from an almost null
geodesic in real coordinates which bounces off the singularity at a fixed boundary time $t_c$. This would imply
that in the CFT correlation function, there is a light cone singularity at $t=t_c$. This is problematic because
such a singularity is ruled out in the CFT on very generic grounds. However, it was shown in \cite{fidkowski} that
the CFT correlation function is in fact dominated by a complexified geodesic. There is a branch cut in the CFT
correlation function at $t=t_c$, and the information about the black hole singularity is contained in the analytic structure near $t=t_c$.

\subsection{Where are the microstates of the black hole? }
\label{mathur}
An important breakthrough in our understanding of black holes was the realization that the horizon area of black
holes has all the properties of thermodynamical entropy \cite{Bek}. This seemed to suggest that there exists a large
number of microstates (of the order of the exponential of the horizon area in appropriate units) building up the
black hole. In some settings, this large number of states can be reproduced from the dual holographic field
theory. However, the question remains how all these different states are manifested in terms of the actual gravity
description of the black hole.

In a recent series of papers \cite{Mathur} , it was suggested that in fact the black hole is not one classical solution
having a singularity and a horizon, but rather is a "coarse grained" description of an ensemble of different
geometries, each being completely regular, and each corresponding to a microstate in the dual field theory. These
geometries are all very similar to each other and to the 'naive' black hole geometry, when probed with particles
of large wavelength, but differ in a small region which defines the location of a 'horizon'. One must note that
this horizon has nothing to do with the classical horizon we are used to. it is not a special surface and there is
no singularity inside it. It is just the characteristic location where all the different geometries start to
differ from each other.

Such geometries were actually built for a specific system of branes - the two charge rotating D1-D5 system
(characterized by the charges of the branes: $Q_1$ and $Q_5$ respectively and by an angular momentum related
parameter $a$), which is supersymmetric (1/4 BPS) \cite{LMM}. At some scaling limit, it describes $AdS_3\times S^3$, whose
holographic dual is a 1+1 dimensional CFT with $SO(4)$ symmetry . For this system the geometries built are
asymptotically flat (at $r>>(Q_1Q_5)^{1/4}$), and have a finite throat, which at some length scales
($(Q_1Q_5)^{1/4}>>r>>a$) describes an $AdS_3\times S^3$ geometry, and deeper ($a>r$) describes a different
geometry, whose details depend on the specific state associated. These systems are also related by U-duality to
the 1/4 BPS supertubes \cite{supertubes} and for critical values of the angular momentum have a similar 'blow-up'
mechanism \cite{LMM} .

One problematic aspect of these systems is that their macroscopic entropy really vanishes, as the number of
microscopic states, although finite, is too small to give any macroscopic entropy. It would be nice if these ideas could be also shown for
the3-charge D1-D5-momentum system (which is 1/8 BPS), where the macroscopic entropy is nonzero. The main problem is
that it is not known how to build a general family of geometries dual to all these microstates. First attempt in
this direction have recently been made in \cite{Lunin,MathurP}.

Another approach to counting the geometrical microstates of this system is to try and manipulate the expression
for the elliptic genus of the conformal field theory and rewrite it as a sum over different geometries with
$AdS_3\times S^3$ asymptotics \cite{JdB, Farey}.

\section{Chern-Simons theory}

Pure Gravity in three dimensions has many special and interesting features, making it, on the one hand a
convenient laboratory for studying gravity and holography, but on the other hand less generic and harder to
generalize to a different number of dimensions. Many of these features are related to the fact that three
dimensional gravity can be rewritten as a topological Chern-Simons theory \cite{WittenCS} \footnote{This is also true for three dimensional supergravity \cite{HMS} .}.

Starting with the regular Einstein action
\begin{equation}
\label{Eaction}
{ S = {1\over 16\pi G}\int_M d^3x \sqrt{|g|}[R-2\Lambda]},
\end{equation}
one can change variables from the metric $g_{\mu\nu}$ to the first order forms - the dreibeins $e_\mu^a$ (such
that $g_{\mu\nu}=e^a_\mu e^a_\nu$) and the spin connections $\omega^a_\mu \equiv {1\over
2}\epsilon^a_{\;\;bc}\omega^{bc}_\mu$, where the action is re-written as
\begin{equation}
\label{rewrit}
{S={2\over 16\pi G}\int_M d^3x
[e^a\wedge (d\omega^a+{1\over 2}\epsilon_{abc}\omega^b\wedge \omega^c)+{\Lambda\over 6}e^a\wedge e^b\wedge e^c]}.
\end{equation}
Then changing variables from $e^a,\omega^a$ to ${\cal A}^a_{L,R}=\omega^a\pm{1\over \sqrt{-\Lambda}}e^a$ ,the
action becomes
\begin{eqnarray}
S &=&k_LS_{CS}[{\cal A}_L]-k_RS_{CS}[{\cal A}_R],\nonumber \\ S_{CS}[{\cal A}] &\equiv&
{1\over 2}\int_M Tr({\cal A}\wedge d{\cal A} +{2\over 3}{\cal A}\wedge {\cal A}\wedge {\cal A}),
\label{cs}
\end{eqnarray}
 and $k_{L}=
{\sqrt{-\Lambda}\over 8\pi G}\,,\,k_R=k_L^*$. In \ref{cs} we regard ${\cal A}_{L,R}$ as taking value in a Lie algebra.
In case we are discussing Lorentzian gravity they are two independent 1-forms, each taking values in $sl(2,R)$,
and in case we are discussing Euclidean gravity, they are complex conjugates of each other, taking values in
$sl(2,C)$.

The fact that one can recast pure gravity as a gauge theory which is topological is very appealing. However, the
change of variables we presented here between the second and first order formalisms is naive and ignores an
important subtlety. Namely, the mapping is not one-to-one, and it is not clear whether one should include also
degenerate ${\cal A}_{L,R}$ or not. Therefore there might be problems defining the measure in the path integrals
for the gravity action and for the Chern-Simons theory. Another possibly related problem is that it seems the
Chern-Simons theory cannot account for the black hole entropy (as calculated from its horizon area), and predicts
a much smaller number of states \cite{Ent}. One therefore might question whether the theories actually exist.
There are also other interesting subtleties related to the AdS/CFT correspondence in this setting, which we would not get
into here (see for example \cite{gukovetal}).

One interesting aside is that in fact many 3-manifolds admit a hyperbolic structure (i.e. admit metrics of
constant negative curvature). For such compact manifolds, there exists a natural complexification of the volume of
the manifold, which involves the Chern-Simons topological invariant, and has good analytic properties \cite{Thurston} :
\begin{equation}
\label{ZofM}
{Z(M) \sim \exp [ {2\over \pi}Vol(M) +4\pi i\ CS(M)]},
\end{equation}
(where above we set $\Lambda=-1$). It is not clear
if and what would be the importance and interpretation of this invariant in the context of the AdS/CFT
correspondence. For some discussion of this invariant and its possible applications see \cite{Gukov}.

\section{Time dependence}
There are many interesting issues that arise in time dependent or cosmological space times. For example, there are generically multiple natural vacua that we can choose in such backgrounds. Also, most such backgrounds exhibit cosmological particle production. The holographic theory reflects these phenomena in an interesting way. We now discuss an explicit example of such a background.

\subsection{AdS bubbles of nothing}
By starting with the AdS Schwarzchild solution and performing a double analytic continuation, we obtain
interesting time dependent backgrounds which are called AdS bubbles of nothing \cite{bubble1,vijaysimon,bubble2}.
The AdS Schwarzchild metric is given in (\ref{adsblackhole}).  The analytic continuation $t\rightarrow i \chi$ and
$ \theta \rightarrow i \tau$ yields a time dependent space time which is a vacuum solution to five dimensional
gravity with a negative cosmological constant: \be ds^2=(1+{r^2 \over l^2}-{2m \over r})d \chi^2 +(1+{r^2 \over
l^2}-{2m \over r})^{-1} dr^2 +r^2(-d \tau^2 + \cosh^2 \tau d\phi^2). \ee This is a smooth spacetime if $\chi$ is
periodic with period ${4\pi r_+ l^2  \over 3r_+^2+l^2}$. Here $r_+$ is the minimum value of the coordinate $r$ and is the largest positive root of the equation $r_+^3+l^2 r_+-2ml^2=0$.
 For a fixed $\tau$ and for
$r>r_+$, the space is basically $S^1_\chi \times S^1_\phi$. As $r \rightarrow r_+$, the circle parameterized by
$\chi$ collapses and the circle parameterized by $\phi$ approaches a finite size $r_+^2 \cosh^2 \tau$. This circle
is the boundary of a bubble of nothing. The metric on the boundary of the bubble is 2d de Sitter space. This space
is asymptotically AdS with the conformal boundary being two dimensional de Sitter space times a circle. So the
holographic dual to this bubble lives on dS${}_2 \times S^1$ \cite{vijaysimon}.

Similar time dependent spacetimes can be constructed by performing double analytic continuations of AdS-Kerr and
Reissner-Nordstrom AdS black holes.

\section{Zero and positive cosmological constant}

In the previous sections solutions of the Einstein equations with
a negative cosmological constant and in particular Anti-de Sitter
space played a prominent role. It is an obvious question to what
extent similar results can be obtained for spaces with zero or a
positive cosmological constant. Much less is known in these cases,
and it is in fact not clear to what extent a meaningful
holographic duality can be formulated.

\subsection{Positive cosmological constant}

The maximally symmetric solution of the Einstein equations with a
positive cosmological constant is de Sitter space. In Euclidean
signature it is simply a $d$-sphere, whereas in Lorentzian
signature it is the time-dependent geometry
\be \label{j1}
ds^2 = -dt^2 + \cosh^2 t d\Omega_{d-1}^2,
\ee
with $d\Omega_{d-1}^2$ the metric on a round $(d-1)$-sphere. The
metric (\ref{j1}) describes a sphere that contracts exponentially
in the past until it reaches a fixed size, and then expands
exponentially again. According to recent experimental data, the
present day expanding universe is well described by de Sitter
space.

Whether or not quantum gravity (or rather, string theory) on a space like (\ref{j1}) is dual to a field theory of
some sort is unclear. Attempts to find such field theories run into various kinds of problems (see e.g.
\cite{dsdual}). Unfortunately, despite a lot of recent work, a clean explicit example of a solution to the string
theory equations of motion of the form (\ref{j1}) is still lacking. Such a solution would obviously be very
helpful in exploring the physics of quantum gravity in de Sitter space.

The space (\ref{j1}) has two boundaries and a cosmological horizon. Associated to the cosmological horizon is a
finite Hawking temperature, and in addition it has a finite area, similar to what one has for a black hole
horizon. This suggests that there might exist some version of holography which applies to cosmological horizons
and associates a finite entropy to them. If one additionally believes that a version of black hole complementarity
applies to cosmological horizons - i.e. the Hilbert space of a single observer is sufficient to describe both
sides of the horizon - then a dual description of de Sitter space might involve a theory with a finite dimensional
Hilbert space. In such a theory there is not enough resolving power to measure arbitrary small distances, and
therefore it can at best yield a dual of (\ref{j1}) which is a good description for a finite amount of time but
not asymptotically as $t \rightarrow \pm \infty$.

If we try to apply the AdS/CFT philosophy more directly to
(\ref{j1}), we should first find all possible solution of the
Einstein equations that are asymptotically identical to
(\ref{j1}). It is known \cite{friedrich} that for sufficiently
small deformations of the boundary metrics a smooth solution with
the same asymptotic behavior still exists. However, the situation
for large deformations has not been resolved. Under significantly
large perturbations de Sitter space can break into pieces and in
particular disconnected geometries (for example so-called big
bang/crunch geometries) will start to contribute
\cite{dsbreak,pm1}. Therefore it is quite possible that the sum
over geometries involved in a putative dS/CFT correspondence will
involve a much larger set of metrics and geometries than just
(\ref{j1}).

Perhaps a better strategy is to first study some simpler aspects
of de Sitter space before engaging in a full-fledged holographic
correspondence. For example, whether or not a positive mass
theorem for de Sitter space exists and if so what its precise
formulation is, is still an open problem. A preliminary positive
mass theorem was described in \cite{pm1}, and it was verified in
many different examples, but a closer look \cite{pm2} suggests
that its formulation is not quite complete as it stands.

\subsection{Zero cosmological constant}

The case of zero cosmological constant is at least as problematic as the case of a positive cosmological constant.
The maximally symmetric solution of the Einstein equations is Minkowski space (or smooth quotients thereof). There
have been a few attempts at finding a dual description of quantum gravity in Minkowski space. First, one can try
to find a dual description of a subset of Minkowski space by putting suitable ``holographic screens'' in it, see
\cite{bou}. This has not led to a concrete dual description however. Another approach involves taking a
decompactification limit of AdS/CFT \cite{pol}. This turns out to be quite difficult and has not led to a precise
dual description either.

A third approach involves the conformal boundary of Minkowski
space, in particular past and future null infinity. It has been
known for a long time that the asymptotic symmetry group of this
boundary is the so-called BMS group, an infinite dimensional
group. In the spirit of AdS/CFT, one might try to look for a
theory that carries representations of this large group. This is
also quite problematic, see \cite{gio} for a recent discussion.

A final approach involves slicing of Minkowski space in Anti-de
Sitter and de Sitter slices. These slices are given by the
equation $\eta_{\mu\nu} x^{\mu} x^{\nu} = r$, where $r<0$ gives
rise to AdS slices and $r>0$ gives rise to dS slices. The case
$r=0$ corresponds to the light-cone. The idea is now to apply holography to each
slice separately and then to combine the results. In this way one
obtains a holographic dual theory that lives on the boundary of
the light-cone, i.e. in two dimensions less, and which has
infinitely many degrees of freedom  \footnote{
One problem is that the dS/AdS slicing of Minkowski space doesn't really survive small deformations. 
In 3+1 dimensions, there are no deformations which give
self-similar slices (i.e slices which only differ from each other by rescaling). Such deformations may exist in higher dimensions,
however the global spaces always have a singularity. We thank M. Anderson for his comment regarding this. See \cite{anderson} for more details.}. Although various miracles
happen \cite{solo} it remains to be seen whether these have
essentially a kinematic origin, or whether they reveal some true
holographic nature of Minkowski space.


\begin{thebibliography}{99}


\bibitem{ads1}
J.~M.~Maldacena,
Adv.\ Theor.\ Math.\ Phys.\  {\bf 2}, 231 (1998) [Int.\ J.\
Theor.\ Phys.\  {\bf 38}, 1113 (1999)] [arXiv:hep-th/9711200].

\bibitem{ads2}
S.~S.~Gubser, I.~R.~Klebanov and A.~M.~Polyakov,
Phys.\ Lett.\ B {\bf 428}, 105 (1998) [arXiv:hep-th/9802109].

\bibitem{ads3}
E.~Witten,
Adv.\ Theor.\ Math.\ Phys.\  {\bf 2}, 253 (1998)
[arXiv:hep-th/9802150].

\bibitem{ads4}
O.~Aharony, S.~S.~Gubser, J.~M.~Maldacena, H.~Ooguri and Y.~Oz,
Phys.\ Rept.\  {\bf 323}, 183 (2000) [arXiv:hep-th/9905111].


\bibitem{Farey}{R.~Dijkgraaf, J.~M.~Maldacena, G.~W.~Moore and E.~Verlinde, ``A black hole farey tail,''
[arXiv:hep-th/0005003].}

\bibitem{anderson}
M.~T.~Anderson,
``Geometric aspects of the AdS/CFT correspondence,''
arXiv:hep-th/0403087.

\bibitem{phasetrans}
E.~Witten,
``Anti-de Sitter space, thermal phase transition, and confinement in  gauge
Adv.\ Theor.\ Math.\ Phys.\  {\bf 2}, 505 (1998)
[arXiv:hep-th/9803131].

\bibitem{hawkingpage}
S.~W.~Hawking and D.~N.~Page,
Commun.\ Math.\ Phys.\  {\bf 87}, 577 (1983).

\bibitem{gukovetal}
S.~Gukov, E.~Martinec, G.~Moore and A.~Strominger,
``Chern-Simons gauge theory and the AdS(3)/CFT(2) correspondence,''
arXiv:hep-th/0403225.


\bibitem{hawking}
S.~W.~Hawking, C.~J.~Hunter and D.~N.~Page,
``Nut charge, anti-de Sitter space and entropy,''
Phys.\ Rev.\ D {\bf 59}, 044033 (1999)
[arXiv:hep-th/9809035].


\bibitem{AndLor}{M.~T.~Anderson, P.~T.~Chrusciel and E.~Delay,
``Non-trivial, static, geodesically complete vacuum space-times
with a negative cosmological constant,'' Jour. High Energy
Physics, 10 (2002) 063, 1-27. }


\bibitem{Galletal}{G.~J.~Galloway, S.~Surya and E.~Woolgar,
``A uniqueness theorem for the adS soliton,''
Phys.\ Rev.\ Lett.\  {\bf 88} (2002) 101102
[arXiv:hep-th/0108170] ; G.~J.~Galloway, S.~Surya and E.~Woolgar,
``On the geometry and mass of static, asymptotically AdS spacetimes, and
the uniqueness of the AdS soliton,''
Commun.\ Math.\ Phys.\  {\bf 241} (2003) 1
[arXiv:hep-th/0204081].}


\bibitem{HorMyers}{G.~T.~Horowitz and R.~C.~Myers,
``The AdS/CFT correspondence and a new positive energy conjecture for general
relativity,'' Phys.\ Rev.\ D {\bf 59} (1999) 026005
[arXiv:hep-th/9808079].}


\bibitem{SSW}{
S.~Surya, K.~Schleich and D.~M.~Witt,``Phase transitions for flat adS
black holes,'' Phys.\ Rev.\ Lett.\  {\bf 86} (2001) 5231
[arXiv:hep-th/0101134].}


\bibitem{ThDyn}{D.~R.~Brill, J.~Louko and P.~Peldan,
``Thermodynamics of (3+1)-dimensional black holes with toroidal or higher
genus horizons,'' Phys.\ Rev.\ D {\bf 56} (1997) 3600
[arXiv:gr-qc/9705012].}



\bibitem{Vanzo}{L.~Vanzo, ``Black holes with unusual topology,''
Phys.\ Rev.\ D {\bf 56} (1997) 6475 [arXiv:gr-qc/9705004].}


\bibitem{Page}{D.~N.~Page, ``Phase transitions for gauge theories on tori
from the AdS/CFT correspondence,'' arXiv:hep-th/0205001.}

\bibitem{TCT}{G.~J.~Galloway, K.~Schleich, D.~Witt and E.~Woolgar, ``The AdS/CFT correspondence conjecture and
topological censorship,'' Phys.\ Lett.\ B {\bf 505}, 255 (2001) [arXiv:hep-th/9912119].
}


\bibitem{eternal}{J.~M.~Maldacena, ``Eternal black holes in Anti-de-Sitter,'' JHEP {\bf 0304}, 021 (2003)
[arXiv:hep-th/0106112]. }

\bibitem{WY}{E.~Witten and S.~T.~Yau, ``Connectedness of the boundary in the AdS/CFT correspondence,'' Adv.\ Theor.\
Math.\ Phys.\ {\bf 3}, 1635 (1999) [arXiv:hep-th/9910245].}




\bibitem{CG}{M.~l.~Cai and G.~J.~Galloway, ``Boundaries of zero scalar curvature in the AdS/CFT correspondence,''
Adv.\ Theor.\ Math.\ Phys.\  {\bf 3}, 1769 (1999) [arXiv:hep-th/0003046].}

\bibitem{Brett}{B.~McInnes,``Quintessential Maldacena-Maoz cosmologies,''
JHEP {\bf 0404} (2004) 036, [arXiv:hep-th/0403104].\ B.~McInnes, ``Answering a basic objection to Bang/Crunch holography,''
arXiv:hep-th/0407189.}


\bibitem{CutGlue}{S.~Aminneborg, I.~Bengtsson, D.~Brill, S.~Holst and P.~Peldan, ``Black holes and wormholes in 2+1
dimensions,'' Class.\ Quant.\ Grav.\ {\bf 15}, 627 (1998) [arXiv:gr-qc/9707036] .\ S.~Aminneborg, I.~Bengtsson,
S.~Holst and P.~Peldan, ``Making Anti-de Sitter Black Holes,'' Class.\ Quant.\ Grav.\  {\bf 13}, 2707 (1996)
[arXiv:gr-qc/9604005].\ M.~Banados, ``Constant curvature black holes,'' Phys.\ Rev.\ D {\bf 57}, 1068 (1998)
[arXiv:gr-qc/9703040].\ K.~Krasnov, ``Analytic continuation for asymptotically AdS 3D gravity,'' Class.\ Quant.\
Grav.\  {\bf 19}, 2399 (2002) [arXiv:gr-qc/0111049].\ D.~R.~Brill, ``Multi-Black-Hole Geometries in
(2+1)-Dimensional Gravity,'' Phys.\ Rev.\ D {\bf 53}, 4133 (1996) [arXiv:gr-qc/9511022]. }


\bibitem{BHTZ}
M.~Banados, M.~Henneaux, C.~Teitelboim and J.~Zanelli,
``Geometry of the (2+1) black hole,''
Phys.\ Rev.\ D {\bf 48}, 1506 (1993)
[arXiv:gr-qc/9302012].


\bibitem{bns}
V.~Balasubramanian, A.~Naqvi and J.~Simon,
``A multi-boundary AdS orbifold and DLCQ holography: A universal holographic
arXiv:hep-th/0311237.

\bibitem{WHs}{J.~Maldacena and L.~Maoz, ``Wormholes in AdS,'' JHEP {\bf 0402}, 053 (2004) [arXiv:hep-th/0401024].}



\bibitem{Bers}{L.~Bers, ``Simultaneous Uniformization,'' Bull.\ Amer.\ Math.\ Soc.\ {\bf 66}, 94 (1960).\ L.~Bers,
``Spaces of Kleinian groups,'' In Maryland conference in several complex variables I, p. 9-34, Springer-Verlag
lecture notes in Math. No. 155, 1970.}


\bibitem{Krasnov}{K.~Krasnov, ``Holography and Riemann surfaces,'' Adv.\ Theor.\ Math.\ Phys.\ {\bf 4}, 929 (2000)
[arXiv:hep-th/0005106].}


\bibitem{hawkinginfo}
S.~W.~Hawking,
``Breakdown Of Predictability In Gravitational Collapse,''
Phys.\ Rev.\ D {\bf 14}, 2460 (1976).


\bibitem{barbon}
J.~L.~F.~Barbon and E.~Rabinovici,
``Long time scales and eternal black holes,''
Fortsch.\ Phys.\  {\bf 52}, 642 (2004)
[arXiv:hep-th/0403268].


\bibitem{horomalda}
G.~T.~Horowitz and J.~Maldacena,
``The black hole final state,''
JHEP {\bf 0402}, 008 (2004)
[arXiv:hep-th/0310281].

\bibitem{solodukhinA}{D.~Birmingham, I.~Sachs and S.~N.~Solodukhin,
``Relaxation in conformal field theory, Hawking-Page transition, and
quasinormal/normal modes,''
Phys.\ Rev.\ D {\bf 67}, 104026 (2003)
[arXiv:hep-th/0212308].}

\bibitem{solodukhinB}{S.~N.~Solodukhin,``Can black hole relax unitarily?,''
arXiv:hep-th/0406130.}

\bibitem{kleban}
M.~Kleban, M.~Porrati and R.~Rabadan, ``Poincare Recurrences and
Topological Diversity,'' arXiv:hep-th/0407192.

\bibitem{KOS}
P.~Kraus, H.~Ooguri and S.~Shenker,
``Inside the horizon with AdS/CFT,''
Phys.\ Rev.\ D {\bf 67}, 124022 (2003)
[arXiv:hep-th/0212277].
\bibitem{leviross}
T.~S.~Levi and S.~F.~Ross,
``Holography beyond the horizon and cosmic censorship,''
Phys.\ Rev.\ D {\bf 68}, 044005 (2003)
[arXiv:hep-th/0304150].

\bibitem{vijaylevi}
V.~Balasubramanian and T.~S.~Levi,
``Beyond the veil: Inner horizon instability and holography,''
arXiv:hep-th/0405048.
\bibitem{fidkowski}
L.~Fidkowski, V.~Hubeny, M.~Kleban and S.~Shenker,
``The black hole singularity in AdS/CFT,''
JHEP {\bf 0402}, 014 (2004)
[arXiv:hep-th/0306170].


\bibitem{Bek}{J.~D.~Bekenstein, ``Black Holes And Entropy,'' Phys.\ Rev.\ D {\bf 7}, 2333 (1973).\ J.~D.~Bekenstein,
``Generalized Second Law Of Thermodynamics In Black Hole Physics,'' Phys.\ Rev.\ D {\bf 9}, 3292 (1974).}

\bibitem{Mathur}{O.~Lunin and S.~D.~Mathur, ``Metric of the multiply wound rotating string,'' Nucl.\ Phys.\ B {\bf
610}, 49 (2001) [arXiv:hep-th/0105136].\ O.~Lunin and S.~D.~Mathur, ``AdS/CFT duality and the black hole
information paradox,'' Nucl.\ Phys.\ B {\bf 623}, 342 (2002) [arXiv:hep-th/0109154].\ O.~Lunin and S.~D.~Mathur,
``Rotating deformations of AdS(3) x S**3, the orbifold CFT and strings in the pp-wave limit,'' Nucl.\ Phys.\ B
{\bf 642}, 91 (2002) [arXiv:hep-th/0206107].\  O.~Lunin, S.~D.~Mathur and A.~Saxena, ``What is the gravity dual of
a chiral primary?,'' arXiv:hep-th/0211292.}






\bibitem{supertubes}{D.~Mateos and P.~K.~Townsend, ``Supertubes,'' Phys.\ Rev.\ Lett.\  {\bf 87}, 011602 (2001)
[arXiv:hep-th/0103030].\ R.~Emparan, D.~Mateos and P.~K.~Townsend, ``Supergravity supertubes,'' JHEP {\bf 0107}
(2001) 011 [arXiv:hep-th/0106012]\ D.~Mateos, S.~Ng and P.~K.~Townsend, ``Tachyons, supertubes and
brane/anti-brane systems,'' JHEP {\bf 0203}, 016 (2002) [arXiv:hep-th/0112054].\ D.~Mateos, S.~Ng and
P.~K.~Townsend, ``Supercurves,'' Phys.\ Lett.\ B {\bf 538}, 366 (2002) [arXiv:hep-th/0204062].}

\bibitem{LMM}{O.~Lunin, J.~M.~Maldacena and L.~Maoz , ``Gravity Solutions for the D1-D5 System with Angular
Momentum,'', [arXiv:hep-th/0212210].}

\bibitem{Lunin}{O.~Lunin, ``Adding momentum to D1-D5 system,'' [arXiv:hep-th/0404006].}




\bibitem{JdB}{J.~de~Boer, ``Large N Elliptic Genus and AdS/CFT Correspondence,'' JHEP {\bf 9905}, 017 (1999)
[arXiv:hep-th/9812240].}


\bibitem{WittenCS}{E.~Witten, ``(2+1)-Dimensional Gravity As An Exactly Soluble System,'' Nucl.\ Phys.\ B {\bf 311},
46 (1988).}

\bibitem{HMS}{M.~Henneaux, L.~Maoz and A.~Schwimmer, ``Asymptotic Dynamics and Asymptotic Symmetries of
Three-Dimensional Extended AdS Supergravity,'' Ann.\ Phys.\ {\bf 282}, 31 (2000) [arXiv:hep-th/9910013].}

\bibitem{Ent}{D.~Kutasov and N.~Seiberg, ``Number of Degrees Of Freedom, Density Of States And Tachyons In Strong
Theory And Cft,'' Nucl.\ Phys.\ B {\bf 358}, 600 (1991). E.~J.~Martinec, ``Conformal field theory, geometry, and
entropy,'' [arXiv:hep-th/9808021].  S.~Carlip, ``What we don't know about BTZ black hole entropy,'' Class.\
Quant.\ Grav.\ {\bf 15}, 3609 (1998) [arXiv:hep-th/9806026].}

\bibitem{Thurston}{W.~Thurston, ``Three-Dimensional Manifolds, Klienian Groups and Hyperbolic space forms,'' Bull.\
Amer.\ Math.\ Soc.\ (N.S.)\ {bf 6}, 357 (1982). W.~Z.~Neumann and D.~Zagier, Topology {\bf 24}, 307 (1985).
Yoshida, ``The invariant of hyperbolic 3-manifolds,'' Invent.\ Math.\ {\bf 81}, 473 (1985).  }

\bibitem{bubble1}
D.~Birmingham and M.~Rinaldi,
``Bubbles in anti-de Sitter space,''
Phys.\ Lett.\ B {\bf 544}, 316 (2002)
[arXiv:hep-th/0205246].


\bibitem{vijaysimon}
V.~Balasubramanian and S.~F.~Ross,
``The dual of nothing,''
Phys.\ Rev.\ D {\bf 66}, 086002 (2002)
[arXiv:hep-th/0205290].
\bibitem{bubble2}
A.~M.~Ghezelbash and R.~B.~Mann,
``Nutty bubbles,''
JHEP {\bf 0209}, 045 (2002)
[arXiv:hep-th/0207123].

\bibitem{dsdual}
E.~Witten, ``Quantum gravity in de Sitter space,''
arXiv:hep-th/0106109; A.~Strominger, ``The dS/CFT
correspondence,'' JHEP {\bf 0110}, 034 (2001)
[arXiv:hep-th/0106113]; V.~Balasubramanian, J.~de Boer and
D.~Minic, ``Exploring de Sitter space and holography,'' Class.\
Quant.\ Grav.\  {\bf 19}, 5655 (2002) [Annals Phys.\  {\bf 303},
59 (2003)] [arXiv:hep-th/0207245]; M.~Spradlin and A.~Volovich,
``Vacuum states and the S-matrix in dS/CFT,'' Phys.\ Rev.\ D {\bf
65}, 104037 (2002) [arXiv:hep-th/0112223]; N.~Goheer, M.~Kleban
and L.~Susskind, ``The trouble with de Sitter space,'' JHEP {\bf
0307}, 056 (2003) [arXiv:hep-th/0212209]; T.~Banks, ``A critique
of pure string theory: Heterodox opinions of diverse dimensions,''
arXiv:hep-th/0306074.

\bibitem{friedrich}
H.~Friedrich, ``On the existence of $n$-geodesically complete or
future complete solutions of Einstein's field equations with
smooth asymptotic structure,'' Commun.\ Math.\ Phys.\ {\bf 107},
587 (1986).

\bibitem{dsbreak}
P.~H.~Ginsparg and M.~J.~Perry, ``Semiclassical Perdurance Of De
Sitter Space,'' Nucl.\ Phys.\ B {\bf 222}, 245 (1983);
J.~C.~Niemeyer and R.~Bousso, ``The nonlinear evolution of de
Sitter space instabilities,'' Phys.\ Rev.\ D {\bf 62}, 023503
(2000) [arXiv:gr-qc/0004004]; R.~Bousso, O.~DeWolfe and
R.~C.~Myers, ``Unbounded entropy in spacetimes with positive
cosmological constant,'' Found.\ Phys.\  {\bf 33}, 297 (2003)
[arXiv:hep-th/0205080].

\bibitem{pm1} V.~Balasubramanian, J.~de Boer and D.~Minic,
``Mass, entropy and holography in asymptotically de Sitter
spaces,'' Phys.\ Rev.\ D {\bf 65}, 123508 (2002)
[arXiv:hep-th/0110108].

\bibitem{pm2} R.~Clarkson, A.~M.~Ghezelbash and R.~B.~Mann,
``Entropic N-bound and maximal mass conjectures violation in four dimensional Taub-Bolt(NUT)-dS spacetimes,''
Nucl.\ Phys.\ B {\bf 674}, 329 (2003) [arXiv:hep-th/0307059]; M.~Anderson, ``On the structure of asymptotically de
Sitter and anti-de Sitter spaces,'' arXiv:hep-th/0407087.

\bibitem{bou}
R.~Bousso, ``A Covariant Entropy Conjecture,'' JHEP {\bf 9907},
004 (1999) [arXiv:hep-th/9905177].

\bibitem{pol}
J.~Polchinski, ``S-matrices from AdS spacetime,'' arXiv:hep-th/9901076.

\bibitem{gio}
G.~Arcioni and C.~Dappiaggi, ``Exploring the holographic principle
in asymptotically flat spacetimes  via the BMS group,'' Nucl.\
Phys.\ B {\bf 674}, 553 (2003) [arXiv:hep-th/0306142];
``Holography in asymptotically flat space-times and the BMS
group,'' arXiv:hep-th/0312186.

\bibitem{solo}
J.~de Boer and S.~N.~Solodukhin, ``A holographic reduction of
Minkowski space-time,'' Nucl.\ Phys.\ B {\bf 665}, 545 (2003)
[arXiv:hep-th/0303006]; S.~N.~Solodukhin, ``Reconstructing
Minkowski space-time,'' arXiv:hep-th/0405252.

\bibitem{MathurP}
S.~Giusto, S.~D.~Mathur and A.~Saxena, ``Dual geometries for a set of 3-charge microstates,''
arXiv:hep-th/0405017; S.~Giusto, S.~D.~Mathur and A.~Saxena,``3-charge geometries and their CFT duals,''
arXiv:hep-th/0406103.

\bibitem{Gukov}
S.~Gukov, ``Three-dimensional quantum gravity, Chern-Simons theory, and the A-polynomial,''
arXiv:hep-th/0306165.


\end{thebibliography}
\end{document}